\newcommand{\CLASSINPUToutersidemargin}{16.08mm}
\documentclass[conference]{IEEEtran}
\makeatletter
\def\ps@headings{%
\def\@oddhead{\mbox{}\scriptsize\rightmark \hfil \thepage}%
\def\@evenhead{\scriptsize\thepage \hfil \leftmark\mbox{}}%
\def\@oddfoot{}%
\def\@evenfoot{}}
\makeatother
\usepackage{cite}
\usepackage{amsmath}
\usepackage{booktabs}
\usepackage{amsfonts}
\usepackage{subfigure}
\usepackage{algorithm}
\usepackage{algorithmic}
\let\labelindent\relax
\usepackage{enumerate}
\usepackage{enumitem}
\usepackage[utf8]{inputenc}
\usepackage[english]{babel}
\usepackage{amsthm}
\usepackage{url}
\usepackage[pdftex]{graphicx}
\usepackage{comment}
\newtheorem{definition}{Definition}
\newtheorem{theorem}{Theorem}
\graphicspath{{pictures/}} 
\begin{document}

\title{e-SAFE: Secure, Efficient and Forensics-Enabled Access to Implantable Medical Devices}


\author{\IEEEauthorblockN{Haotian Chi\IEEEauthorrefmark{1}, Longfei Wu\IEEEauthorrefmark{2}, Xiaojiang Du\IEEEauthorrefmark{1}, Qiang Zeng\IEEEauthorrefmark{1}, Paul Ratazzi\IEEEauthorrefmark{3}}
\IEEEauthorblockA{\IEEEauthorrefmark{1}Department of Computer and Information Sciences, Temple University, Philadelphia, PA 19122, USA \\
\IEEEauthorrefmark{2}Department of Mathematics and Computer Science, Fayetteville State University, Fayetteville, NC 28301, USA \\
\IEEEauthorrefmark{3}Air Force Research Laboratory, Rome, NY 13440, USA\\
Email: \{htchi, dux, qzeng\}@temple.edu, lwu@uncfsu.edu, edward.ratazzi@us.af.mil}
}
\maketitle

\begin{abstract}
To facilitate monitoring and management, modern Implantable Medical Devices (IMDs) are often equipped with wireless capabilities, which raise the risk of malicious access to IMDs. Although schemes are proposed to secure the IMD access, some issues are still open. First, pre-sharing a long-term key between a patient's IMD and a doctor's programmer is vulnerable since once the doctor's programmer is compromised, all of her patients suffer; establishing a temporary key by leveraging proximity gets rid of pre-shared keys, but as the approach lacks real authentication, it can be exploited by nearby adversaries or through man-in-the-middle attacks. Second, while prolonging the lifetime of IMDs is one of the most important design goals, few schemes explore to lower the communication and computation overhead all at once. Finally, how to safely record the commands issued by doctors for the purpose of forensics, which can be the last measure to protect the patients' rights, is commonly omitted in the existing literature. Motivated by these important yet open problems, we propose an innovative scheme \emph{e-SAFE}, which significantly improves security and safety, reduces the communication overhead and enables IMD-access forensics. We present a novel lightweight compressive sensing based encryption algorithm to encrypt and compress the IMD data simultaneously, reducing the data transmission overhead by over 50\% while ensuring high data confidentiality and usability. Furthermore, we provide a suite of protocols regarding device pairing, dual-factor authentication, and accountability-enabled access. The security analysis and performance evaluation show the validity and efficiency of the proposed scheme.
\end{abstract}



%
\IEEEpeerreviewmaketitle

\section{Introduction}

Implantable Medical Devices (IMDs) are gaining increasing popularity in the medical industry. IMDs are embedded into patients' bodies to monitor their medical conditions by collecting a range of physiological values (e.g., heart rate, blood glucose and neural activity) and to provide specific therapies (e.g., glucose injection or heart beat regulation) \cite{rushanan2014sok}. To make it more convenient to monitor patients' health status and manage IMDs' configuration, many IMDs are equipped with wireless interfaces, enabling wireless communication between IMDs and an external device called \emph{programmer}. A doctor's programmer can access the patient's IMD to read sensor data, and/or modify treatment configurations through the on-board radios, which largely promotes patients' mobility and eliminating surgical procedures.

Unfortunately, most of the existing wireless medical devices lack sufficient security mechanisms to protect patients from malicious attacks. For example, implantable cardiac defibrillators (ICDs) and pacemakers contain a magnetic switch that can be activated by sufficiently strong magnetic field \cite{Medtronic}. Some research works \cite{halperin2008pacemakers, HealthCom2011} have reported software radio attacks against a commercial ICD and an insulin pump, respectively. The attacks can steal the information stored in the IMD and reprogram it to change the prescribed therapy. McAfee's Barnaby Jack showed that he could command the delivery of a deadly 830-volt shock to pacemakers made by several manufacturers at a distance up to 50 feet away \cite{kirk2012pacemaker}. The consequence of malicious access to IMDs can be fatal. Therefore, the access to IMDs should be strictly controlled to prevent the illegal exposure of patient data and the life-threatening modification of its settings.

Due to the high computational complexity and energy consumption, public-key cryptography is usually not suitable for resource-restricted IMDs. Most of the solutions employ light-weight symmetric cryptography based authentication schemes. In this aspect, the most challenging issue is how to generate and distribute the shared key for the secure channel establishment between an IMD and a programmer. Pre-loading the shared key \cite{halperin2008pacemakers, HealthCom2011, Liu} between each pair of the IMD and the programmer, is not practical and may not be secure. Once a programmer is compromised, all the shared keys stored in it may be leaked and all the corresponding IMDs may be attacked by impersonating the programmer. Since one patient can visit multiple doctors, the attack surface is further enlarged.
Some other schemes set up temporary keys in a distributed manner based on proximity. They assume that only the programmer operated by the doctor can get close to the patient, and thus only it can extract the same key materials as the IMD, such as electrocardiograms \cite{CCS2013}, vibration \cite{Vibration}, ultrasound \cite{rasmussen2009proximity}. However, this subclass of access control schemes does not include a real authentication. Instead, they solely rely on the assumption that the adversary is deterred from being physically close to the patient to avoid leaving criminal evidence.

Proxy-based schemes \cite{Cloaker, xu2011imdguard, gollakota2011they} are promising to mitigate the limitations of IMDs' computation and battery capacity by mediating the communication between IMDs and programmers. However, the existing schemes are still vulnerable to various attacks. Attackers can spoof the IMD that the patient is in emergency by jamming the \emph{Cloaker} to acquire open access \cite{Cloaker}. In \emph{IMDGuard} \cite{xu2011imdguard}, the patient's time-varied Electrocardiography (ECG) signals are used to establish the secret key, assuming that ECG signals can only be measured with skin contact. However, it was challenged by literature \cite{rostami2013balancing} claiming that the effective key length can be reduced by a man-in-the-middle attack, and literature \cite{poh2011advancements} demonstrated that ECG can be inferred by recording and analyzing a video that contains the face of a person for a period of time.

By investigating existing schemes, we find most schemes require either a hardware modification on IMDs or programmers to retrieve the key material sources (e.g., ECG, vibration, etc.), or an dedicated proxy device. Moreover, the security and robustness of generating key material based on proximity can be impaired or even invalidated by various attacks. Last but not the least, most, if not all, schemes only focus on how to derive the shared key between the programmer and the IMD (or the proxy), without considering the significant issues about minimizing the power consumption of the energy-critical IMDs and providing forensics \cite{cheng2017lightweight} of critical operations on IMDs by the authorized doctors.

Motivated by the above insights, we propose \emph{e-SAFE}, an efficient Secure-Access and Forensics-Enabled scheme, to ensure the security of IMDs. In e-SAFE, the patient uses a common smartphone rather than a dedicated device to perform a dual-factor authentication on the requesting programmer. Based on the shared key derived in the authentication, we also present protocols for the read and write accesses to IMDs. Any critical write operations performed by a doctor is accountable. To make it more scalable and secure, the programmer in our system does not store any security-related materials of IMDs. We also present a novel efficient and secure data transmission scheme based on compressive sensing (CS) to provide confidentiality on the insecure wireless medical telemetry and considerably reduce the communication overhead, which usually accounts for approximately 70\% power consumption of IMDs.

\vspace{-0mm}
\textbf{Contributions}. We summarize our main contributions as follows.
\vspace{-2mm}
\begin{enumerate}[leftmargin=*]
  \item We propose e-SAFE, which incorporates the common used smartphone as a proxy. A patient's smartphone can be used to conduct dual-factor authentication and access control on the requesting programmer. Accountability and emergency treatment are supported.
  \item We design a novel compressive sensing based encryption algorithm that fits in with our system architecture and considerably reduces the transmission overhead in IMDs without introducing additional computation overhead.
  \item We provide security analyses of the e-SAFE scheme's resistance to various passive and active attacks.
  \item We implement a prototype of the system and conduct performance evaluation to validate the effectiveness and efficiency of the proposed scheme.
\end{enumerate}

\vspace{-1mm}
\section{Preliminaries}
In this section, we present the threat model, the system objectives, and the proposed system architecture.  
\vspace{-2mm}

\subsection{Threat Model}
\vspace{-1mm}
Potential adversaries include \emph{passive} and \emph{active} attackers. Passive attackers simply eavesdrop on the wireless channel to obtain private information; they neither modify the message in the channel, nor communicate with the patient's smartphone or IMD. Besides eavesdropping, active attackers may intercept, tamper with or replay messages in a wireless channel to launch various attacks. The objective of active attacks includes obtaining the patient's private information or sending malicious commands to disturb the normal functions of IMDs. We assume that attackers have regular computational bounds, e.g., they cannot perform efficient brute force attacks to solve the discrete logarithm problem. However, we make no assumption on the attacker's communication capability. They may be able to eavesdrop the wireless channel, block messages, and even hijack the communication between entities in the system. Attacks by exploiting operating system or firmware vulnerabilities \cite{wu2014security}, malware \cite{liang2014permission}, etc., are beyond the scope of this paper.

\vspace{-2.5mm}
\subsection{System Objectives}
\vspace{-0.5mm}

\textbf{Security}: In general, safety, confidentiality, and integrity are considered in a secure system \cite{du2008security}. In terms of safety, IMDs should only be accessible to authorized programmers used by authenticated doctors, and any write access to the IMD should be accountable in case of a medical dispute. As for privacy, the patient's physiological data and treatment data should be kept confidential in wireless transmission. Besides, integrity ensures that critical data cannot be modified by unauthorized entities and its origin could be verified.

\textbf{Low Overhead}: The computation and communication overhead in the patient's devices, especially the IMD, should be minimized to save power consumption.

\textbf{Flexible Architecture}: While a patient is usually equipped with the same IMD for a period of time, he may see different doctors in different clinics, healthcare centers or hospitals. Hence, the scheme should allow an IMD to be accessed by multiple programmers safely.


\vspace{-2.3mm}
\subsection{System Architecture}
\begin{figure}[!ht]
  \vspace{-2mm}
  \centering
  \includegraphics[width=0.32\textwidth]{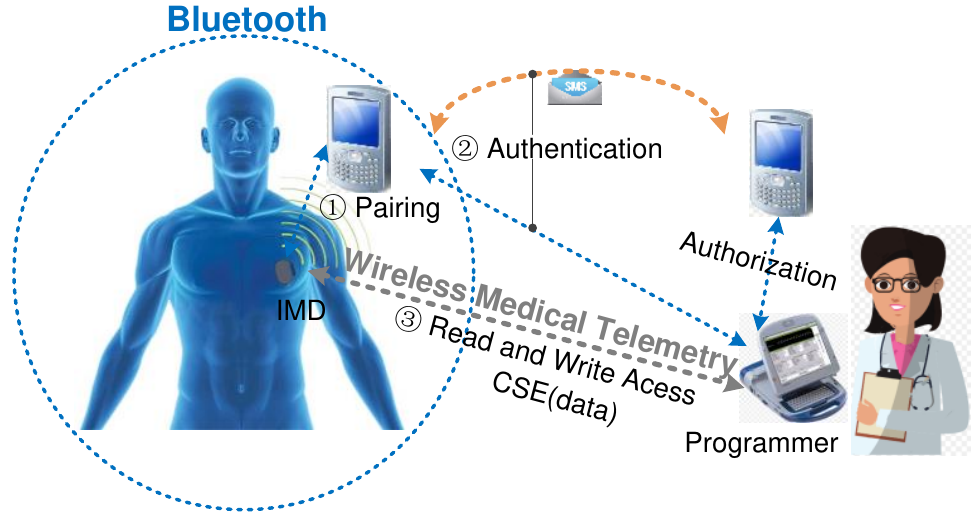}\\
  \caption{System Architecture}\label{fig_system}
  \vspace{-2mm}
\end{figure}

\vspace{-0.9mm}
The system comprises three major entities: IMD, Smartphone and Programmer, as shown in Fig. \ref{fig_system}.

\begin{itemize}[leftmargin=*]
  \item \textbf{IMD}: A patient's IMD is equipped with a Bluetooth module to communicate with his smartphone within a limited distance. The IMD communicates with a doctor's programmer through the conventional wireless medical telemetry.
  \item \textbf{Smartphone}: Every patient and each doctor hold a smartphone. The patient's smartphone communicates with his IMD via Bluetooth, and communicates with a programmer wirelessly, e.g., Bluetooth, WiFi, or cellular networks.
  \item \textbf{Programmer}: The programmer refers to the device used by doctors in medical centers or hospitals. The programmer can read from or write to the patient's IMD. The programmer may also have interfaces for external I/O and storage devices, i.e., a keyboard and a USB flash drive.
  \vspace{-0.5mm}
\end{itemize}

Besides, at the initialization phase in our system, a public key infrastructure (PKI) is used to issue a private/public key pair and a certificate to each doctor. This is only a one-time cost at the initialization phase. After the initialization, our scheme does not need a PKI to verify the identity and public key of a doctor. Furthermore, this kind of basic PKI is common in many e-health applications and becomes an essential security entity with the development of e-health.

\section{Secure and Efficient IMD Data Transmission}
\label{sec_seidt}
In this section, we first present a novel encryption framework based on compressive sensing for secure and efficient IMD data transmission. Then we briefly discuss the potential improvement on data compression by uniform quantification.

\vspace{-2mm}
\subsection{Background: Compressive Sensing}
\vspace{-0.5mm}
\label{sub_sec_comsen}
Compressive sensing \cite{donoho2006compressed, candes2006near} combines traditional sensing and compressing into a single process by exploiting data sparsity. Let $\mathbf x\in \mathbb{R}^{\rm N}$ be a real-valued $N$-dimensional original signal that can be sparsely represented on a certain orthonormal basis. $\mathbf{x}$ is $s$-sparse if $\mathbf{x}$ can be expanded upon a basis $\mathbf{\Psi}\in\mathbb R^{\rm N\times N}$ with no more than $s$ non-zero coefficient entries, i.e., $\mathbf{x} = \mathbf{\Psi b}$, where $\mathbf{b}\in\mathbb{R}^{\rm N}$ is the sparse vector satisfying $||\mathbf{b}||_0 \leq s$. In the sensing process, the original signal vector is multiplied by a sensing matrix $\mathbf{\Phi}\in\mathbb{R}^{\rm M\times N}$ that is incoherent with $\mathbf{\Psi}$, obtaining the $M$-measurements vector $\mathbf{y} = \mathbf{\Phi x}$. As $M \ll N$, the original signal $\mathbf{x}$ is compressed to $\mathbf{y}$, with a compressing rate ${\rm CR} = 100(N-M)/N$. Thus, the compressive sensing process is expressed as:

\vspace{-4mm}

\begin{equation}
\centering
\mathbf{y} = \mathbf{\Phi x = \Phi\Psi b = \Theta b}
\end{equation}
\vspace{-7mm}

To ensure robust and precise recovery from $\mathbf{x}$, $\mathbf{\Phi}$ should satisfy the Restricted Isometry Property (RIP). RIP of order $s$ and level $\delta_{s}\in(0, 1)$ (i.e., $(s, \delta_{s})$-RIP) is satisfied if

\vspace{-3mm}
\begin{equation}
(1-\delta_{s})||\mathbf{b}||_{2} \leq ||\mathbf{\Phi\Psi b}||_{2} \leq (1+\delta_{s})||\mathbf{b}||_{2}
\end{equation}
holds for all $s$-sparse $\mathbf{b}\in\mathbb{R}$. If RIP holds, the sparse vector $\mathbf{b}$ can be estimated by solving the basis pursuit (BP) problem in (\ref{eq_bp}):

\vspace{-6mm}

\begin{equation}
\label{eq_bp}
\min\limits_{\hat{\mathbf{b}}\in\mathbb{R}^{\rm N}}||\hat{\mathbf{b}}||_{1} \qquad \text{subjects to} \qquad \mathbf{y = \Phi\Psi\hat{b}}
\end{equation}
There already exist a number of algorithms to reconstruct the original signal, e.g., orthogonal matching pursuit (OMP) \cite{tropp2007signal}, gradient projection \cite{figueiredo2007gradient}, etc.

\vspace{-1.9mm}
\subsection{Motivation}
\vspace{-1mm}
Implantable and wearable medical devices usually acquire sparse signals that can be sensed by compressive sensing, a promising technique to provide efficient and secure transmission for IMD devices due to the following features. First, it can provide a high compression rate since its sampling rate is lower than the Nyquist rate, which is 2 times of the signal's bandwidth. Second, the computation overhead for sensing is much lower than that for signal reconstruction. Besides, compressive sensing can provide data encryption by viewing the sensing matrix $\mathbf{\Phi}$ as a secret key, without which it is computationally difficult to recover $\mathbf{x}$ from $\mathbf{y=\Phi x}$ \cite{orsdemir2008security, rachlin2008secrecy}. Therefore, the IMD can combine data compression and encryption into a single operation by a CS-based algorithm.

However, pre-sharing $\mathbf{\Phi}$ between each IMD and each of its accessing programmers is neither flexible nor safe. A straightforward alternative in our architecture may be that the smartphone generates and distributes a $\mathbf{\Phi}$ to the IMD and the programmer in each session. Nevertheless, it is not practical since receiving such a large matrix causes too much communication overhead for IMDs. To solve the above problems, we propose a novel CS-based encryption algorithm (\emph{CSE}) which fits well with the proxy-based system architecture.

\vspace{-2mm}
\subsection{CS-Based Encryption Algorithm}
\label{sec_cse}
\vspace{-1mm}
In our setting, both the sparse matrix $\mathbf{\Psi}$ and the sensing matrix $\mathbf{\Phi}$ are public. Before presenting the $\emph{CSE}$ algorithm, we first introduce the related operations in Definitions \ref{def_shift}-\ref{def_entry_product}.

\begin{definition}{(Shifting Addition).}
\label{def_shift}
The shifting addition of two random variables $v\in(L_{1}, L_{2})\subset\mathbb{R}$ and $w\in\mathbb{R}$ outputs $(u, \lambda) \leftarrow v\mathbf{\triangleleft\oplus} w$, such that $u = (v + w - L_{1})$ mod $(L_{2} - L_{1}) + L_{1}$, where $L_{1}, L_{2}\in\mathbb{Z}$. $\lambda\in(0, 1)$ is set to 0 if $(v + w)\in [L_{1}, L_{2})$; otherwise, $\lambda = 1$.
\end{definition}

\begin{definition}{(Vectorial Shifting Addition).}
\label{def_vec_shift}
If all entries in $\mathbf{v}$ are \emph{i.i.d.} and have a finite range, the vectorial shifting addition of two vectors $\mathbf{v}, \mathbf{w}\in\mathbb{R}^{\rm N\times 1}$ can be computed by
$(\mathbf u, \mathbf\Lambda) \leftarrow \mathbf{v}\mathbf{\triangleleft\oplus}\mathbf{w}$.
 Let $v_{i}, w_{i}, u_{i}, \lambda_{i}, i = 1, \cdots, N$ be entries of $\mathbf{v, w, u}\in\mathbb{R}^{\rm N\times 1}$, $\mathbf{\Lambda}\in (0, 1)^{\rm N\times 1}$, respectively, and $(u_{i}, \lambda_{i}) \leftarrow v_{i}\mathbf{\triangleleft\oplus}w_{i}$.
\end{definition}

\begin{definition}{(Vectorial Entry-wise Product).}
\label{def_entry_product}
The entry-wise product of two same-sized vectors $\mathbf{v, w}\in\mathbb{R}^{\rm N\times 1}$ is defined as $\mathbf{v}.*\mathbf w = (v_{1}*w_{1}, \cdots, v_{N}*w_{N})$, where $v_{i}, w_{i}, i=1, \cdots, N$ are entries of $\mathbf{v, w}$, respectively.
\vspace{-3mm}
\end{definition}

Based on the above operations, the \emph {CSE} algorithm is defined in the following.

\begin{definition}{(CSE Algorithm).}
An energy-efficient CS-Based encryption scheme on the original signal $\mathbf{x}$ whose entries are bounded in $(L_{1}, L_{2})$, where $L_{1}, L_{2}\in\mathbb{Z}$, is defined as $\rm CSE = (CSGen, CSEnc, CSDec)$ such that,
\end{definition}

\vspace{-2mm}
\textbf{$K_d \leftarrow CSGen(sd, k)$}: is a simple random generation algorithm run by the smartphone, which utilizes a pseudo-random function (PRF) seeded by $sd$ to generate $k$ random $d_i\in[-\lceil range/2\rceil, \lceil range/2\rceil]\subset\mathbb{Z}$, where $i = 1, 2, \cdots, k$ and $range = L_{2} - L_{1}$, then $\mathbf{K_d} = (d_1, \cdots, d_k)^{\mathbf T}$.

\textbf{($\mathbf{y'}, \mathbf{\Lambda}) \leftarrow  CSEnc(\mathbf{K_d}, \mathbf{x})$}: is some simple vector and matrix operations run by the IMD to encrypt the patient data. It takes as input a secret key $\mathbf{K_d}$, the original data $\mathbf{x}$ and also the public $\mathbf{\Phi}$, and outputs the compressed and encrypted data $\mathbf{y'}$, i.e., $(\mathbf p, \mathbf\Lambda) \leftarrow (\mathbf{x} \mathbf{\triangleleft\oplus} \mathbf{K_{d}})$; $\mathbf{y'} = \mathbf{\Phi}\mathbf{\mathbf p}$.

\textbf{$\hat{\mathbf{x}} \leftarrow CSDec(\mathbf{K_d}, \mathbf y', \mathbf\Lambda)$}: includes a vector operation and a reconstruction algorithm typically for compressive sensing, performed by the programmer. It takes as input $\mathbf{K_d}$, $\mathbf{y'}$, $\mathbf{\Lambda}$ and $\mathbf\Phi$, and outputs the estimation $\hat{\mathbf{x}}$ of the original data $\mathbf{x}$. In the first step, it gets the real compressed data $\mathbf{y}$ from $\mathbf{y'}$ by (\ref{eq_cs_dec}), where $\mathbf{sign} = (d_{1}/|d_{1}|, \cdots, d_{N}/|d_{N}|)$ and represents the shifting direction.

\vspace{-7mm}
\begin{equation}
\label{eq_cs_dec}
\mathbf{y} = \mathbf\Phi \mathbf x =  \mathbf{y'} + \mathbf{\Phi} ((L_{2} - L_{1})\mathbf{\Lambda}.*\mathbf{sign} - \mathbf{K_{d}})
\vspace{-3mm}
\end{equation}

After getting $\mathbf y$, an estimation $\mathbf{\hat{x}}$ of $\mathbf{x}$ can be derived by a reconstruction algorithm that solves the problem in (\ref{eq_bp}).

\vspace{-2mm}
\subsection{Further Optimization: Uniform Quantification}
\vspace{-1mm}
Typically, the output of $CSEnc$ is finite precision floating values since the original IMD data $\mathbf{x}$ is a natural signal. To further compress the data transmission, we simply quantify the output $\mathbf{y'}$ with a quantification step $qs$, obtaining $\mathbf{y''}$. Then there are $m = range/qs$ values for the elements in the output of $CSEnc$. The communication overhead in transmission of IMD data can thus be lowered. Moreover, if $m$ is small enough, it is possible for IMDs to encode $\mathbf{y'}$ after quantification by searching a pre-stored small-size code table, e.g., Huffman coding table. We do not discuss encoding in details since it is beyond the scope of this paper, but it can significantly reduce the communication cost.

\vspace{-1mm}
\section{e-SAFE Scheme}
\label{sec_acp}
In this section, we will present the details of our e-SAFE scheme, including four parts: 1) system initialization; 2) device pairing; 3) dual-factor authentication; 4) authorization and access. Headings $R_{j}, j = 1, 2, \cdots, 14$ are used to denote different request types in our protocols.
The notations used in this paper and the correspondence between $j$ and request types are given in Table \ref{table_notations} and \ref{table_request}, respectively.

\begin{table}
\small
\renewcommand{\arraystretch}{1.1}
\caption{Notations}
\label{table_notations}
\centering
\begin{tabular}{ll}
\toprule
Symbol & Description\\
\midrule
$\rm \mathbf{S, I, D, DS}$           & A patient's smartphone, IMD, a doctor's \\& programmer, phone, resp. \\
$\rm R_{j}$           & Message headings \\
$\rm ID_{p}$   & The identity of $p \in \{\mathbf{S, I, D, DS}\}$ \\
$\rm SN_{0}, SN$         & Session numbers \\
$\rm TS_{i}$       & Timestamps, $i = 1, 2, \cdots, 8$ \\
$\rm kdf(\cdot)$         & Key derivation function\cite{kaliski2000pkcs} \\
$\rm Sig(key, msg)$   & RSA signature function \\
$\rm HMAC(key, msg)$     & SHA1-HMAC function\\
$\rm AESEnc(key, msg)$   & AES encryption function\\
$\rm RSAEnc(key, msg)$   & RSA encryption function\\
$\rm CSEnc(key, msg)$   & CS-based encryption function\\

\bottomrule
\end{tabular}
\end{table}

\begin{table}
\small
\caption{Request Types}
\label{table_request}
\centering
\begin{tabular*}{0.47\textwidth}{@{\extracolsep{\fill}} cl|rl|cl}
\toprule
$\mathbf{j}$ & Type & $\mathbf{j}$ & Type & $\mathbf{j}$ & Type\\
\midrule
$1$ & pair\_req & $6$ & read\_allow & $11$  & write\_ready\\
$2$&  pair\_succ & $7$ & read\_ready & $12$  & write\_req\\
$3$ & auth\_req & $8$ & read\_req & $13$  & set\_allow\\
$4$ & optional & $9$ & write\_auth\_req & $14$  & write\_succ\\
$5$ & read\_auth\_req & $10$ & write\_allow \\
\bottomrule
\end{tabular*}
\vspace{-4mm}
\end{table}

\vspace{-1.5mm}
\subsection{System Initialization}
\vspace{-0.5mm}
  Each IMD has a hardcoded master key $K$ and an identity $ID_I$ issued by the device manufacturer and included in the product specification uniquely associated with the device.

  The programmer pre-stores nothing before a doctor loads some information into it by inserting a secure USB device. A doctor's dedicated USB device stores her identity $ID_D$ and private/public key pair $SK$ and $PK$ together with the corresponding public-key certificate $Cert_{PK}$, issued by a trusted PKI whose public key is $PK_{CA}$. With the certificate, the legitimacy of the doctor's public key can be verified.

  The patient's smartphone with an identify $ID_S$ needs to pre-store the CA's public key, the doctor's identity and office phone number or email address. This is reasonable and practical since the patient usually needs to contact and make an appointment with the doctor, and the doctor's office phone number or email address is publicly available. The doctor's phone $ID_{DS}$ does not pre-store any information.

\vspace{-1mm}
\subsection{Device Pairing}
\vspace{-1mm}
\label{subsubsec_pair}
In our scheme, the patient uses his smartphone to perform authentication and authorization when an external programmer requests to access his IMD. To this end, the patient's smartphone and IMD should pair with each other to establish a secure channel. The existing pairing schemes in the Bluetooth protocol stack cannot satisfy our pairing requirement. The IMD is not supposed to know any information about the requesting smartphone, since the pairing process may happen when enrolling a new smartphone if the previously paired one is not available, e.g., it is broken, stolen, etc. Moreover, the patient cannot physically control the IMD to select the target smartphone after the IMD is implanted.

\begin{table}[!ht]
\centering
\renewcommand\arraystretch{1.5}
\caption{Device Pairing Process}
\label{table_pairing}
\begin{tabular*}{0.45\textwidth}{@{\extracolsep{\fill}}ll}
\toprule
(i) $\mathbf{S \rightarrow I}$ & ${\rm R_{1}} || \rm SN_{0} || \rm ID_{S} || \rm TS_{1} || \rm HMAC_{1}$ \\
                     & $ K_{i} = {\rm kdf}(K ||\rm ID_{S} || ID_{I})$ \\
                     & ${\rm HMAC_{1}} = {\rm HMAC}(K_{i}, \rm[R_{1} || SN_{0} || ID_{S} || TS_{1}])$ \\

(ii) $\mathbf{I \rightarrow S}$ & $\rm R_{2} || SN_{0}$ \\
\bottomrule
\end{tabular*}
\vspace{-2mm}
\end{table}

Therefore, we design an application-layer device pairing protocol, as shown in Table \ref{table_pairing}. The patient manually enters into his smartphone the master key $K$ to generate a shared key $K_{i}$, and then sends a pairing request to his IMD as in (i). After checking $\rm HMAC_{1}$ with the key generated in the same way, the IMD notifies the success of pairing. Thereafter, the smartphone and IMD use $K_{i}$ to establish a secure channel.

\vspace{-1mm}
\subsection{Dual-Factor Authentication}
\vspace{-0.5mm}
The patient's smartphone needs to authenticate the programmer held by a doctor before authorizing the programmer to access. We propose a two-factor authentication that verifies both the legality of the doctor and her programmer. The authentication approach is described in Table \ref{table_authentication}.

\vspace{-0mm}
\begin{table}[!ht]
\centering
\renewcommand\arraystretch{1.5}
\caption{Authentication Process}
\label{table_authentication}
\begin{tabular*}{0.47\textwidth}{@{\extracolsep{\fill}}ll}
\toprule
(iii) $\mathbf{D \rightarrow S}$ & ${\rm R_{3} || SN || ID_{D} || TS_{2} || PK || Cert_{PK}}$ \\
(iv) $\mathbf{S \rightarrow D}$ & ${\rm SN || ID_{S} || RSAEnc(PK}, [nonce])$ \\
(v) $\mathbf{S \rightarrow DS}$ & $RM$ \\
(vi) $\mathbf{D \rightarrow S}$ & $\rm R_{4} || SN || TS_{3} || HMAC_{2}$ \\
                     & $K_{p} = {\rm kdf}(RM || {\rm SN} || nonce)$ \\
                     & ${\rm HMAC_{2} = HMAC}(K_{p}, \rm [R_{4} || SN || ID_{D} || ID_{S} || TS_{3}])$ \\

\bottomrule
\end{tabular*}
\vspace{0mm}
\end{table}

 The programmer firstly sends to the patient's smartphone the doctor's public key $PK$ and certificate $Cert_{PK}$ to initiate an authentication process. The patient's smartphone verifies the legitimacy of $PK$ by checking $Cert_{PK}$, and then sends a nonce encrypted with the public key $PK$. Besides, the patient's smartphone sends a random message $RM$ to the doctor's phone via a secure message service, e.g., SMS or email. Once receiving $RM$, the doctor enters it into her programmer manually; hence human involvement is integrated into authentication. Note that only the doctor's programmer and the patient's smartphone can derive a key $K_p$ from $RM$ and $nonce$. The smartphone can then authenticate the programmer by checking $\rm HMAC_{2}$ after receiving it.

 (vi) is shown here for presenting the mechanism of the dual-factor authentication. It is always overridden by the first step of a read process to initiate an access request, which is discussed in Section \ref{subsec_access}.


\vspace{-1mm}
\subsection{Authorization and Access}
\label{subsec_access}
The programmer is authorized to request accesses to the IMD when it shares $K_{p}$ with the patient's smartphone. The programmer may need a read access to retrieve physiological data \emph{data} from the IMD, or sometimes a write access to send a command \emph{CMD} to change the therapy configuration of the IMD based on the analysis of the obtained data. The respective protocols are presented in this subsection.

\subsubsection{Read Process}
\label{subsec_read}
After authenticated, the doctor's programmer initiates a read access process by running the protocol illustrated in Table \ref{table_read}. In this protocol, the programmer first requests a read access authentication by sending a message including $\rm HMAC_{3}$ to the patient's smartphone in (vi). The smartphone then allows the IMD to accept the programmer's read request and distributes the \emph{CSE} key $\mathbf{K_{d}}$ and the derived symmetric key $K_{r}$ to the IMD in (vii). After the IMD returns ``read\_ready", the smartphone forwards the IMD's status and distributes $\mathbf{K_{d}}$ to the programmer in (ix). When receiving the programmer's read request in (x), the IMD sends the CSEnc-ed patient data $C_{2}$ to the programmer. The plaintext of patient data can be obtained by running $\rm CSDec(\cdot)$ on $C_{2}$.

\begin{table}
\centering
\renewcommand\arraystretch{1.4}
\caption{Read Access Process}
\label{table_read}
\begin{tabular*}{0.47\textwidth}{@{\extracolsep{\fill}}ll}
\toprule
(vi) $\mathbf{D \rightarrow S}$ & $\rm R_{5} || SN || TS_{4} || HMAC_{3}$ \\
                     & ${\rm HMAC_{3} = HMAC}(K_{p}, \rm [R_{5} || SN || TS_{4}])$ \\
(vii) $\mathbf{S \rightarrow I}$ & ${\rm R_{6} || SN || ID_{D} || C_{1}} || \rm TS_{5} || HMAC_{4}$ \\
                     & $K_{r} = {\rm kdf}(\mathbf{K_{d}} ||\rm SN)$ \\
                     & ${\rm C_{1} = AESEnc}(K_{i}, [\mathbf{K_{d}}, K_{r}])$ \\
                     & ${\rm HMAC_{4} = HMAC}(K_{i}, \rm[R_{6} || SN || ID_{D} || C_{1} ||TS_{5}])$ \\
(viii) $\mathbf{I \rightarrow S}$ & $\rm R_{7} || SN$ \\
(ix) $\mathbf{S \rightarrow D}$ & ${\rm R_{7} || SN || AESEnc}(K_{p}, [\mathbf{K_{d}}])$ \\
(x) $\mathbf{D \rightarrow I}$ & $\rm R_{8} || SN$ \\
(xi) $\mathbf{I \rightarrow D}$ & $\rm SN || C_{2} || HMAC_{5}$ \\
                     & ${\rm C_{2} = CSEnc}(\mathbf{K_{d}}, data)$ \\
                     & ${\rm HMAC_{5} = HMAC}(K_{r}, \rm [SN || C_{2}])$ \\

\bottomrule
\end{tabular*}
\vspace{-4mm}
\end{table}

\vspace{-0mm}
\subsubsection{Write Process}
\label{subsec_write}
According to the patient's physiological data, the doctor may need to change the treatment configuration by sending a command \emph{CMD} to the patient's IMD, which is a critical operation that can affect the patient's health and probably even life. Our protocol incorporates a signature-based forensics method for the patient's smartphone to record the doctor's operations in case of medical disputes. The write access process is shown in Table \ref{table_write}. When requesting for write access authentication, the programmer sends to the patient's smartphone an RSA signature of information that describes an event ``a doctor $ID_{D}$ sends a command \emph{CMD} to an IMD $ID_{I}$ at $TS_{6}$ based on the data obtained from $C_{2}$ with $\mathbf{K_{d}}$", as shown in (xii). The IMD is activated to accept write access in (xiii). After being notified of the IMD's ``write\_ready" status in (xv), the programmer sends \emph{CMD} to the IMD. (xvii) and (xviii) are executed for forensics purpose. When receiving \emph{CMD} and $C_{2}$ from the IMD, the smartphone can verify the validity of $Sig$ and retain a tuple $\langle{\rm ID_{D}, ID_{S}, ID_{I}}, \mathbf{K_{d}}, {\rm C_{2}, CMD, TS_{6}}, Sig\rangle$ as the evidence of the doctor's write operation. Then, the smartphone allows the IMD to change its configuration according to \emph{CMD} in (xviii).

\begin{table}[!ht]
\centering
\renewcommand\arraystretch{1.3}
\caption{Write Access Process}
\label{table_write}
\begin{tabular}{ll}
\toprule
(xii) $\mathbf{D \rightarrow S}$ & ${\rm R_{9} || SN} || Sig || \rm TS_{6} || HMAC_{6}$ \\
                     & $Sig = {\rm Sig}(SK, {\rm [ID_{D} || ID_{S} || ID_{I} || C_{2} || \mathbf{K_{d}}} || \rm CMD || \rm TS_{6}])$ \\
                     & ${\rm HMAC_{6} = HMAC}(K_{p}, {\rm [R_{9} || SN ||} Sig ||\rm TS_{6}])$ \\
(xiii) $\mathbf{S \rightarrow I}$ & ${\rm R_{10} || SN || ID_{D}} ||\rm TS_{7} || HMAC_{7}$ \\
                     & ${\rm HMAC_{7} = HMAC}(K_{i}, \rm [R_{10} || SN || ID_{D} || TS_{7}])$ \\
(xiv) $\mathbf{I \rightarrow S}$ & $\rm R_{11} || SN$ \\
(xv) $\mathbf{S \rightarrow D}$ & ${\rm R_{11} || SN }$ \\
(xvi) $\mathbf{D \rightarrow I}$ & $\rm R_{12} || SN || C_{3}|| \rm HMAC_{8}$\\
                      & ${\rm C_{3} = AESEnc}(K_{r}, [\rm CMD])$ \\
                     & ${\rm HMAC_{8} = HMAC}(K_{r}, \rm [R_{12} || SN || C_{3}])$ \\

(xvii) $\mathbf{I \rightarrow S}$ & ${\rm SN || C_{2} || AESEnc}(K_{i}, [\rm CMD ||\rm ID_{I}] ||TS_{7})$ \\
(xviii) $\mathbf{S \rightarrow I}$ & $\rm R_{13} || SN || TS_{8} || HMAC_{9}$ \\
                     & ${\rm HMAC_{9} = HMAC}(K_{i}, \rm [R_{13} || SN || TS_{8}])$ \\
(xix) $\mathbf{I \rightarrow D}$ & $\rm R_{14} || SN$ \\

\bottomrule
\end{tabular}
\vspace{-2mm}
\end{table}

\section{Security and Safety Analysis}

\subsection{Resistance to Passive Attacks}
\label{subsec_passive_attack}
To avoid being disclosed to wireless eavesdroppers, all to-be-transmitted sensitive data are protected in our scheme, including the security-related parameters (i.e., $nonce$, $\mathbf{K_d}$ and $K_r$) and \emph{CMD} that may also reveal a patient's health condition. The patient's physiological data are protected by $CSE$. Since $\mathbf{K_d}$ is dynamically updated in different sessions, chosen-plaintext attack (CPA) does not motivate the attacker. We thus consider a ciphertext-only attack (COA) model when analyzing the secrecy of \emph{CSE}.

\emph{CSE} obfuscates the original signal $\mathbf{x}$ by introducing a pre-measurement \emph{i.i.d.} uniformly random noise ($\mathbf{K_{d}}$). We show that the distribution of $\mathbf{x\mathbf{\triangleleft\oplus} K_{d}}$ is statistically close to $\mathbf{O} \mathbf{\triangleleft\oplus} \mathbf{K_{d}}$, where $\mathbf{O}\in\mathbb{R}^{\rm N\times 1}$ is a zero vector.

\vspace{0mm}
\begin{theorem}
The noised signal $\mathbf{x\mathbf{\triangleleft\oplus}K_{d}}$ and a noised zero signal $\mathbf{O} \mathbf{\triangleleft\oplus} \mathbf{K_{d}}$ are statistically indistinguishable.
\vspace{-1mm}
\end{theorem}
\begin{IEEEproof}
Regarding each entry in $\mathbf{x\mathbf{\triangleleft\oplus}K_{d}}$ and $\mathbf{O} \mathbf{\triangleleft\oplus} \mathbf{K_{d}}$, we have two hypotheses $\mathcal{H}_{0}$ and $\mathcal{H}_{1}$, whose ranges are both $[-L_{1}, L_{2})$. The optimal distinguishing strategy is not better than a random guess from (0, 1). Consider the correlation between entries in $\mathbf{x}$, the distinguisher's success probability is $P_{r} = \frac{1}{2} + \mu(\mathbf x)$. The negligible function $\mu(\mathbf x) \rightarrow 0$ because each entry in $\mathbf{x}$ has a random cyclic shift within $[L_{1}, L_{2})$. Therefore, $\mathbf{x\mathbf{\triangleleft\oplus}K_{d}}$ and $\mathbf{O} \mathbf{\triangleleft\oplus} \mathbf{K_{d}}$ are statistically indistinguishable.
\end{IEEEproof}

By making the distribution of $\mathbf x$ close to that of $\mathbf{K_{d}}$, we break $\mathbf x$'s sparsity, which is the underlying basis of reconstruction. Now consider the effect of noise $\mathbf{K_{d}}$ quantitatively when applying compressive sensing. Literature \cite{arias2011noise} presents a detailed proof that the pre-measurement zero-mean white noise (i.e., $\mathbf{K_{d}}$) with covariance $\sigma_{0}^{2}\mathbf{I}$ degrades the signal-to-noise ratio (SNR) by a factor of $\gamma = N/M$ in compressive sensing, which is called \emph{noise folding}. Based on this conclusion, the reconstruction from $\mathbf{y = \Phi(x + K_{d})}$ approximates roughly to solving the additive noise model $\mathbf{y = \Phi x + z}$, where $\mathbf z$ is a white noise with variance $\sigma^{2} = \gamma\sigma_{0}^{2}$.


Hence, our \emph{CSEnc}, at the worst case where few entries are shifted, is equivalent to adding a $\gamma$ larger white noise to the CS measurement of $\mathbf{x}$. In this case, the reconstruction error bound for noised measurement has been widely studied in the compressive sensing domain. Suppose the public $\mathbf{\Phi, \Psi}$ satisfy the 2$s$-th RIP condition, according to Wang et al.'s work \cite{wang2017recovery}, the error bound gain $\Delta E_{\rm CSE}$ by applying $\mathbf{K_{d}}$ is,

\vspace{-3mm}
\begin{equation}
\Delta E_{\rm CSE} \leq \frac{4\sqrt{1+\delta_{2s}}}{1-\delta_{2s}-\sqrt{2}\delta_{2s}}\sigma^{2}
\end{equation}
where $\delta_{2s}<\sqrt{2}-1$. Assume that $\delta_{2s} = 0.3$, we have $\Delta E_{\rm CSE} \leq 16.54\sigma^{2}$. Note that the variance of the original signal $\mathbf x$ is even much smaller than $\sigma_{0}$. The high error gain guarantees CSEnc-ed data irrecoverability without $\mathbf{K_d}$.

\vspace{-1mm}
\subsection{Resistance to Active Attacks}

\vspace{-0mm}
\subsubsection{Impersonation Attack}
To impersonate a legitimate doctor, the attacker needs to have both the doctor's public/private key pairs $PK$ and $SK$ stored in a USB drive and the doctor's phone to pass the dual-factor authentication. In the worst case, the attacker gains both a doctor's USB lost drive and smartphone; theoretically, the attacker can impersonate a real doctor to gain access to a patient's IMD. Our scheme mitigates this threat through human involvement in the system design, preventing the IMDs from being access covertly; the attacker would need to meet the patient physically in the hospital or clinic, which makes this attack impractical.

\vspace{-1mm}
\subsubsection{Replay Attack}

Our scheme can defend the replay attacks. First, we include a session number and timestamp in critical steps that are vulnerable to replay attacks. Second, all the processes are enforced to be single-threaded, i.e., every device only responds to the request it expects in the current session. Last but not least, HMAC checks based on dynamic keys $K_p, K_r$ can also defend against replay attacks.

\subsubsection{Man-in-the-Middle Attack}
\begin{figure}[!ht]
  \centering
  \includegraphics[width=0.33\textwidth]{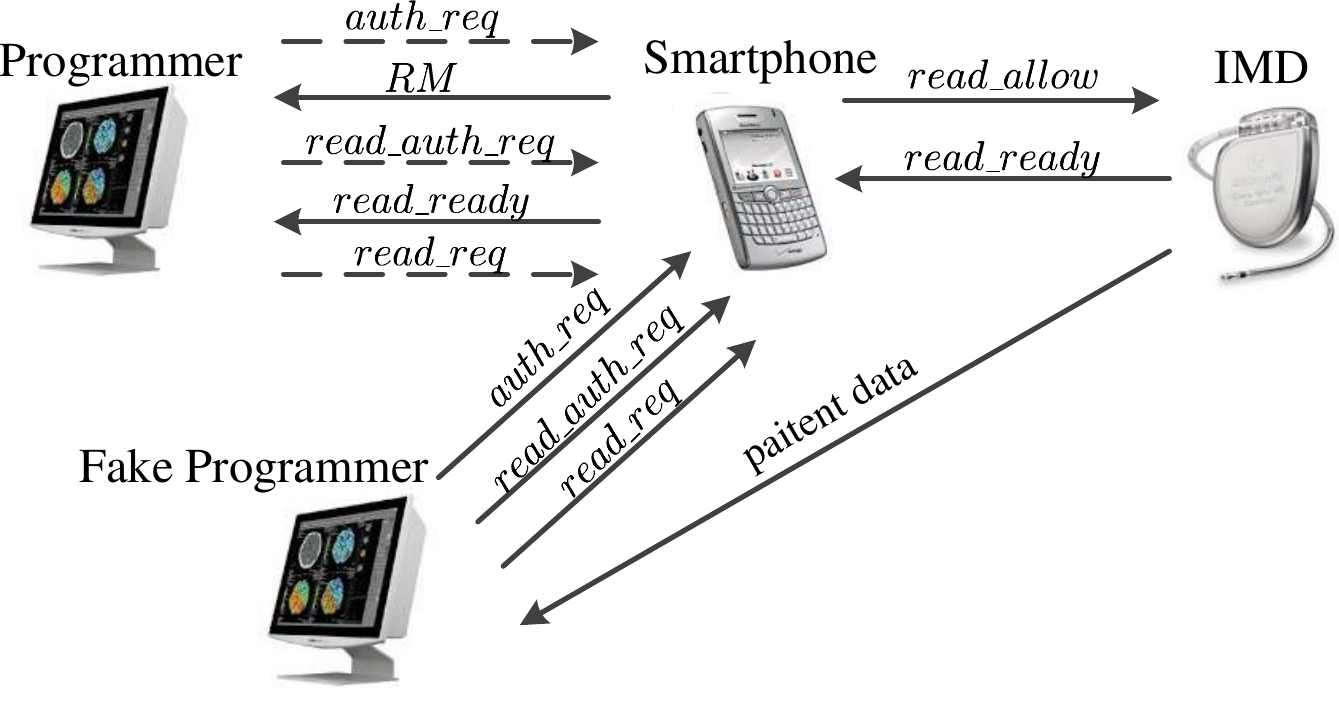}\\
  \caption{Man-in-the-middle attack in read process} \label{fig_man1}
  \vspace{-1mm}
\end{figure}

A man-in-the-middle attack can be conducted if an attacker is able to intercept all the messages in the communication links.
A possible man-in-the-middle attack targeted in revealing patient data is shown in Fig. \ref{fig_man1}. The fake programmer manipulated by the attacker blocks the outgoing communication links of the doctor's programmer and impersonates the real programmer to send the blocked messages. However, the fake programmer can only obtain the ciphertext $\rm C_{2}$ instead of $data$. The attacker can generate neither the secret key $K_p$ without $RM$ nor $K_i$ without IMD's master key $K$, thus cannot get $\mathbf{K_d}$. Consequently, the attacker cannot reconstruct $data$ which is protected by the \emph{CSE}. Likewise, the fake programmer cannot obtain or modify the doctor's command \emph{CMD} without the keys.

\vspace{-2mm}
\subsection{Emergency Open Access}

Since human intervention is required in the authentication, the emergency situation should be carefully considered. That is, even when a patient is unable to assist the authentication, e.g., he goes into a coma, any programmer from the first emergency responder can still access the IMD to give him emergency treatment. Our scheme can deal with this case: the doctor uses the patient's smartphone to send a random message to her own phone to complete the authentication protocol. We recommend that the patient uses his fingerprint to unlock the smartphone as fingerprints have been widely used and the responder can use the patient's fingerprint to unlock his phone even if he is in a comma. Moreover, the device pairing enables a new smartphone to be paired with the IMD, in case that the previously paired smartphone is lost. The extreme condition that the patient's smartphone is lost when he becomes unconscious is beyond the scope of this paper.

\vspace{-1mm}
\section{Performance Evaluation}

\subsection{Experimental Settings}
\vspace{-1mm}
The main challenge in IMD experiments is that the source codes and open platforms from commercial vendors are not available. We choose TelosB with TinyOS 2.1 and Rasberry Pi version 3 Model B with Ubuntu 14.04 LTS system, which are open research platforms for resource constrained systems, as a replacement of the IMD and the programmer, respectively, and implement the prototype of our proposed scheme.
We skip the SMS/email test on smartphone since it is widely used in mobile applications. We use a Sandisk USB drive to store the doctor's identity, public/private keys and certificate.

We use the ECG data records from MIT-BIH Arrhythmia Database \cite{goldberger2000physiobank} to validate the proposed data compression and privacy scheme. Each ECG data record in the database is resampled as a 2-second signal at 256 Hz. We select $\mathbf{\Phi}$ from an independent identically distribution (i.i.d) Gaussian distribution $\mathcal{N}(0, 1/\rm N)$ sampling and exploit a Discrete Fourier Transformation (DFT) basis as the sparse matrix $\mathbf{\Psi}$.

\subsection{Evaluation Results}
\label{sub_sec_eva_result}
In this subsection, we present the evaluation results. We employ the widely used \emph{percentage root-mean square difference (PRD)} \cite{jalaleddine1990ecg} which quantifies the percent error between the original signal vector $\mathbf{x}$ and the recovered $\mathbf{\hat{x}}$, to measure the quality of the recovered signal. PRD also serves as an important security metric in our paper, indicating the secrecy degree that the \emph{CSE} achieves. PRD is defined as

\vspace{-3mm}
\begin{equation}
PRD = \frac{||\mathbf x - \hat{\mathbf x}||_2}{||\mathbf x||_2} \times 100
\end{equation}
Zigel \emph{et al.}'s work \cite{zigel2000weighted} classifies the different values of PRD based on signal quality perceived by a specialist. We use their results as a way to evaluate PRD, as is shown in Table \ref{table_prd_class}:

\begin{table}[!hbp]
\small
\caption{PRD and Corresponding Quality}
\label{table_prd_class}
\centering
\begin{tabular*}{0.3\textwidth}{@{\extracolsep{\fill}}c|c}
\toprule
PRD & Recovered Signal Quality\\
\midrule
0 $\sim$ 2\% & "Very good"\\
2 $\sim$ 9\% & "Very good" or "good"\\
$\geq$ 9\% & "Not good"\\
\bottomrule
\end{tabular*}
\end{table}

\begin{figure*}
  \centering
  \subfigure[A view of reconstructed signal on record 230]{
    \label{fig_view} 
    \includegraphics[width=0.32\textwidth]{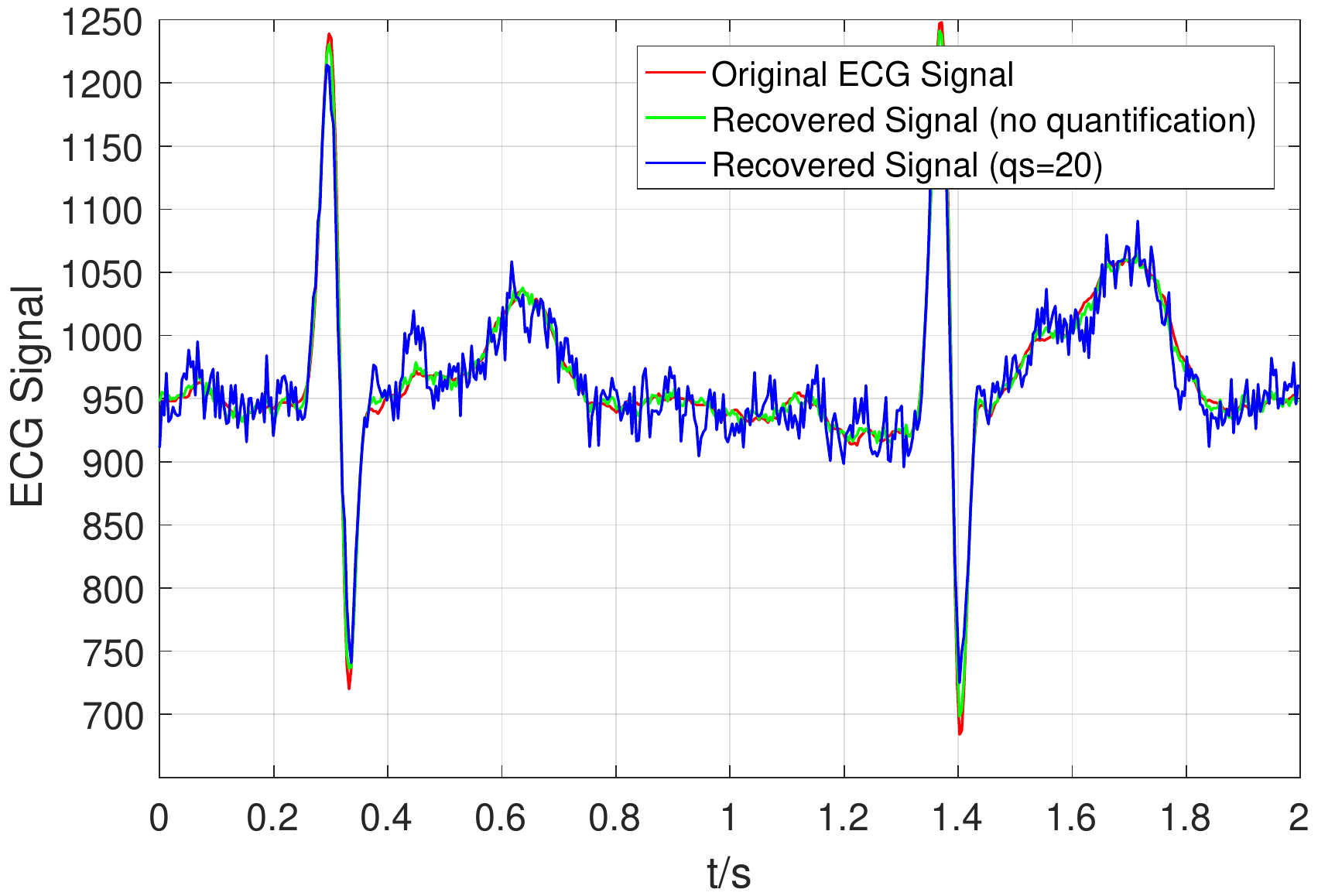}}
  \subfigure[PRD vs. CR under different $qs$]{
    \label{fig_prd} 
    \includegraphics[width=0.30\textwidth]{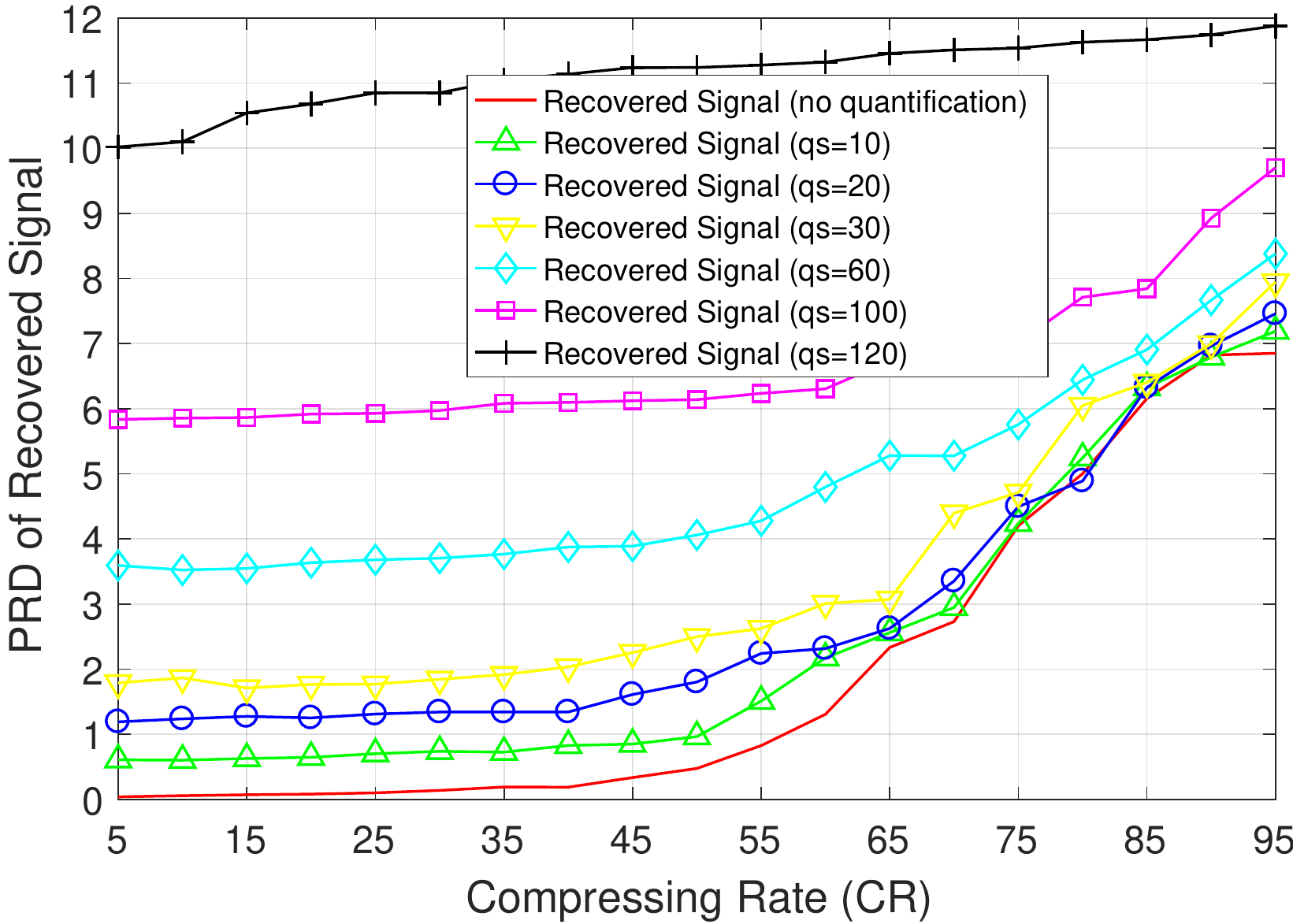}}
  \subfigure[Uniform guess attack with different $\bar{d}$]{
    \label{fig_attack_d} 
    \includegraphics[width=0.315\textwidth]{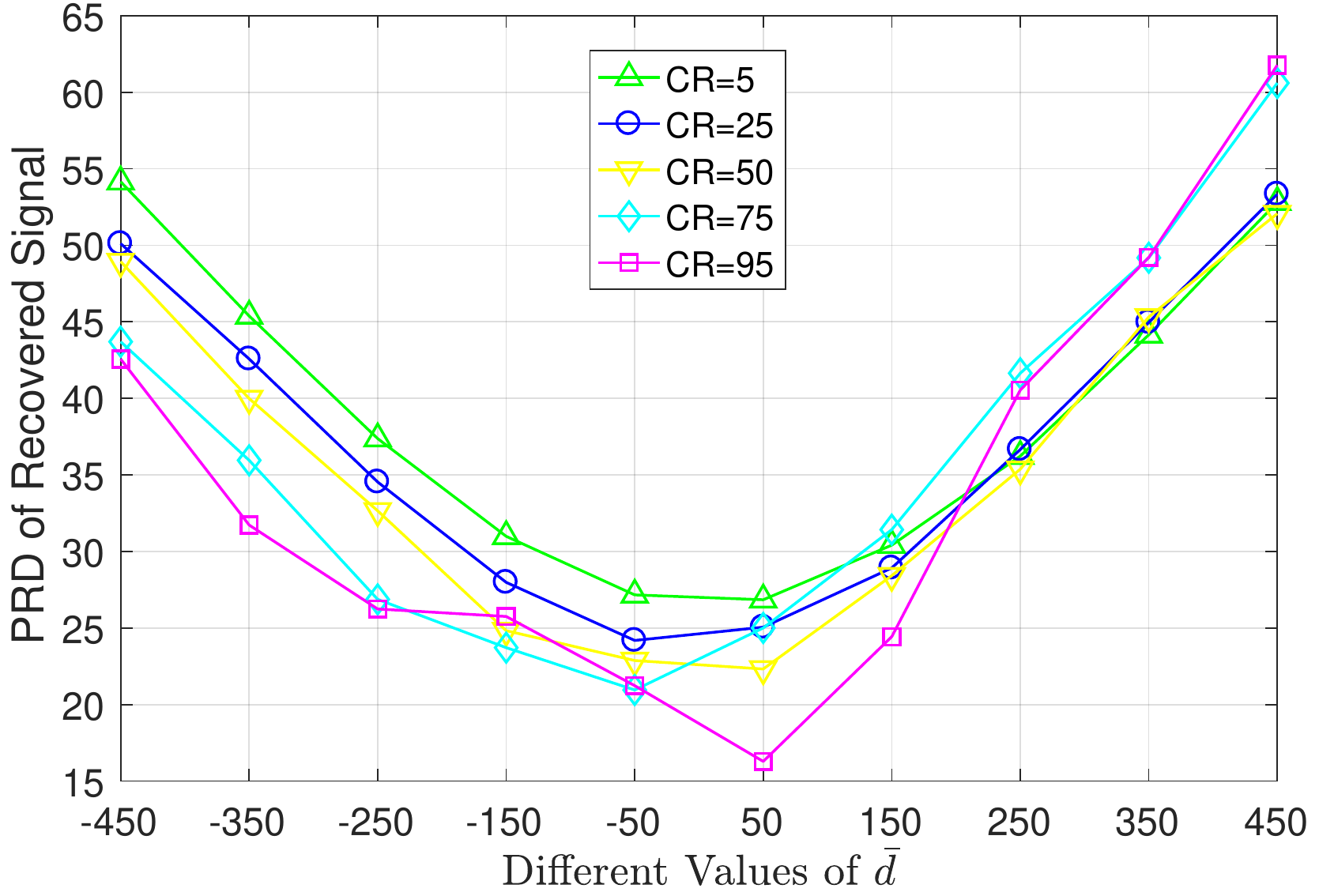}}

  \subfigure[Fragment of record 230 and the recovered signals]{
    \label{fig_attack_view} 
    \includegraphics[width=0.32\textwidth]{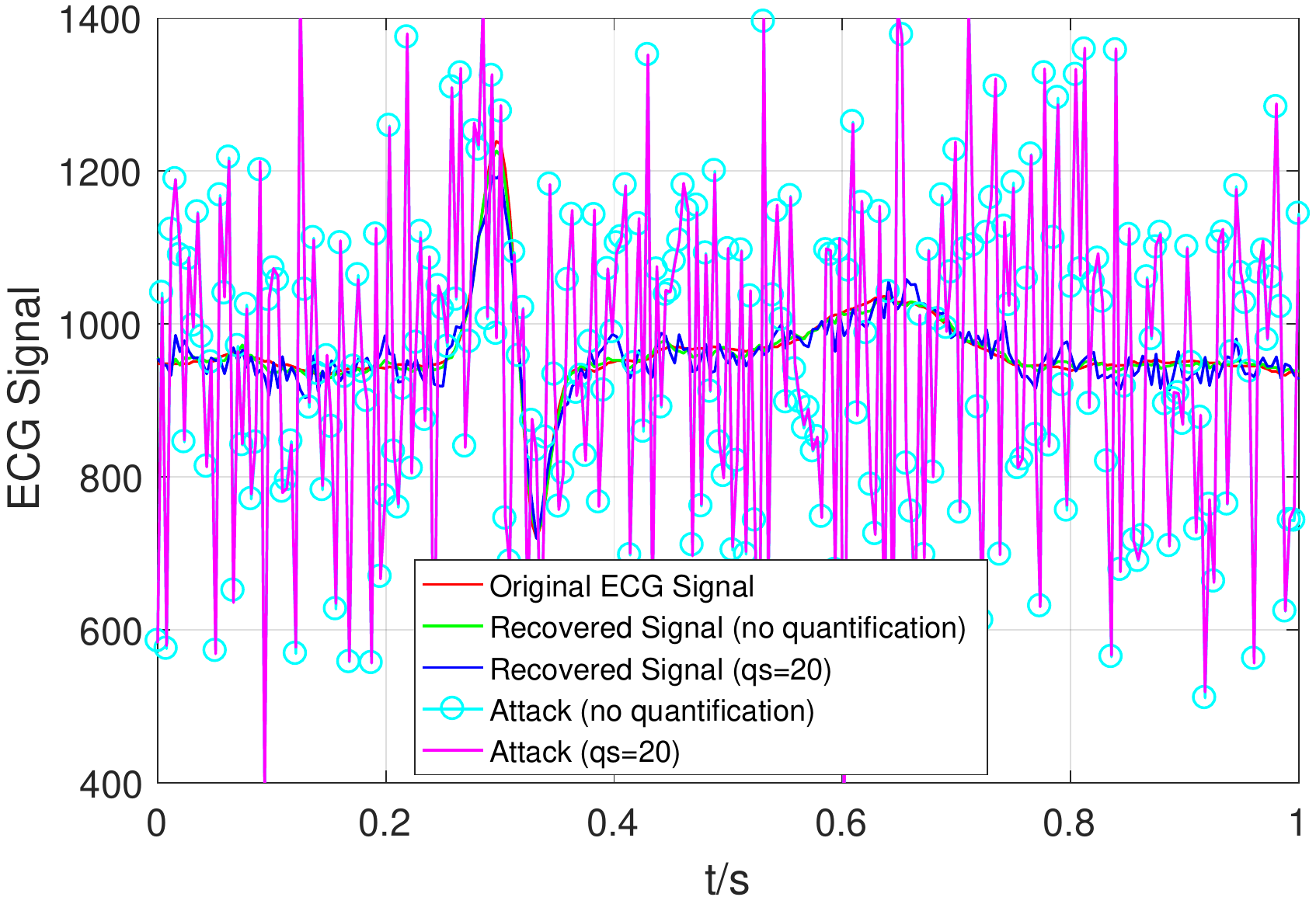}}
  \subfigure[Comparison of legitimate recovery and attack]{
    \label{fig_attack_qs} 
    \includegraphics[width=0.312\textwidth]{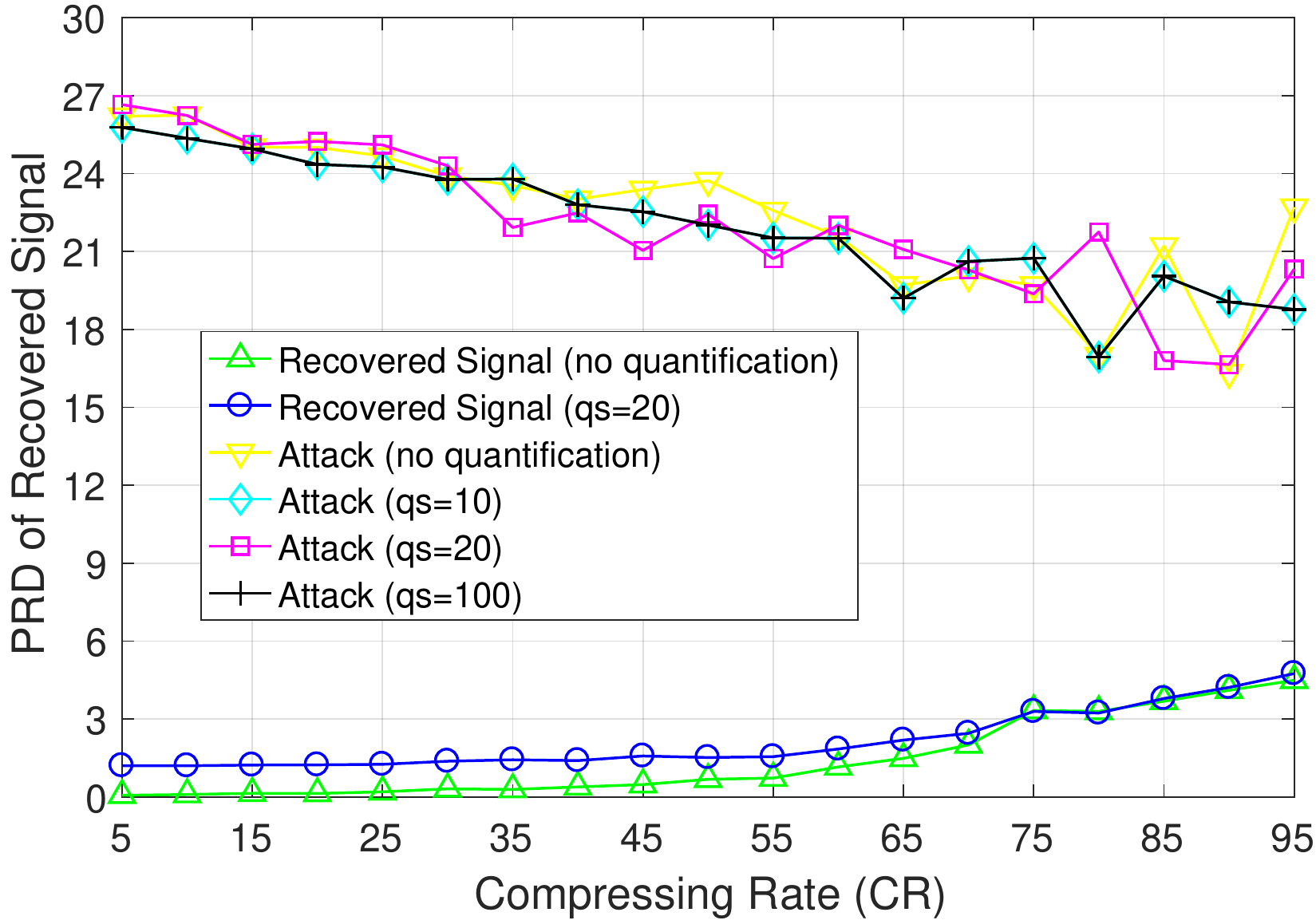}}
  \subfigure[Random guess attack in multiple trials]{
    \label{fig_attack_random} 
    \includegraphics[width=0.32\textwidth]{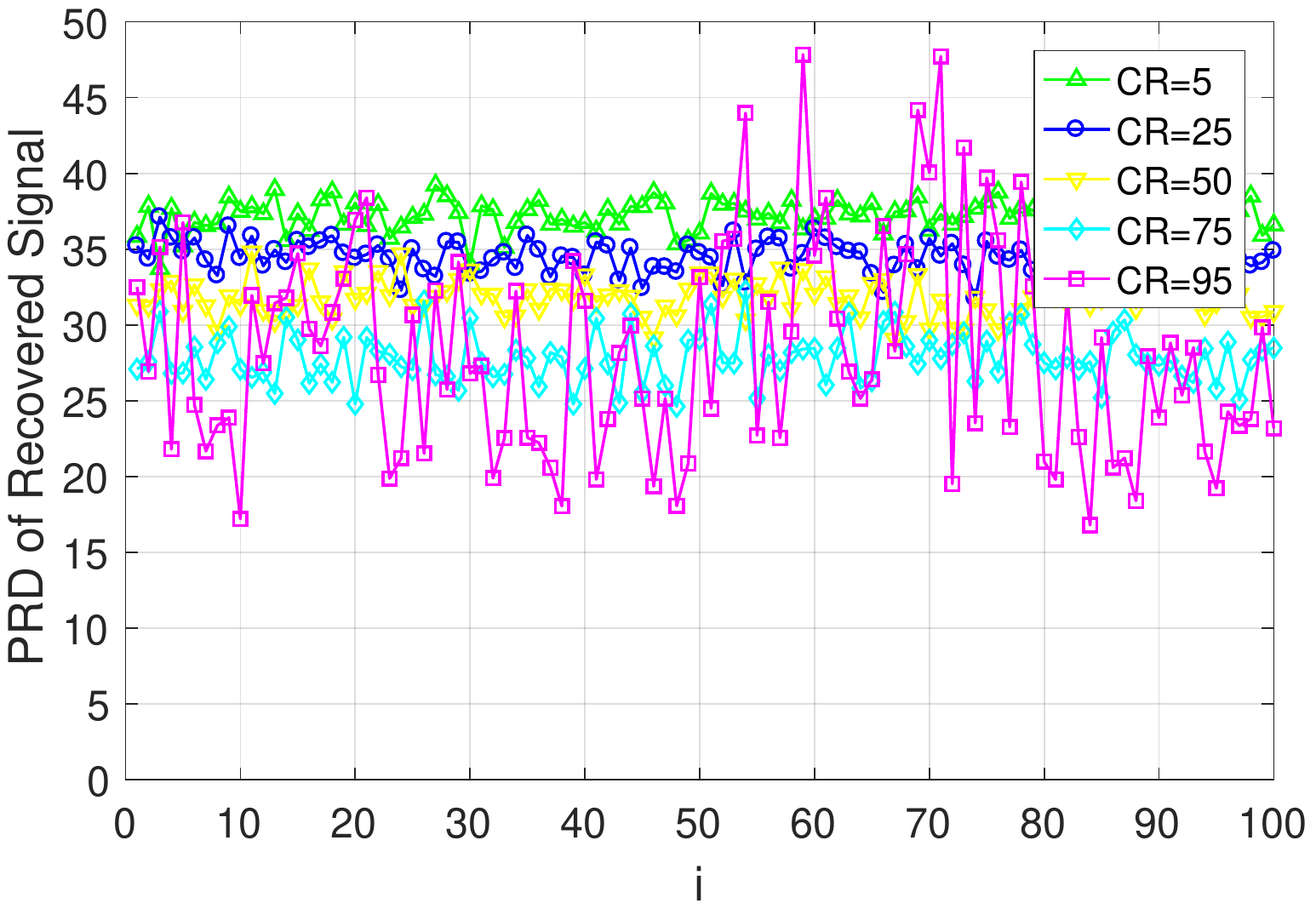}}

  \caption{Usability and secrecy evaluation on data transmission under different situations. }
  \label{fig_secrecy} 
  \vspace{-2.5mm}
\end{figure*}

\subsubsection{Data Usability}
In this paper, the challenge of data usability resides in the reconstruction quality from the compressive sensed and quantified data. We establish an experiment to evaluate the performance of data usability.


We firstly give a view of the snapshot of recovery in Fig. \ref{fig_view} where $\rm CR=50$ and $qs=20$. The original signal can be recovered almost perfectly when no quantification is applied. When the compressed signal $\mathbf{y'}$ is quantified with $qs=20$, the recovery quality degrades slightly, but it's still in the ``very good" class. The result indicates it is effective to apply compressive sensing to sparse IMD data, which requires to be acquired and transmitted accurately enough.


We test the signal reconstruction from compressed signal with quantification step $qs=\langle10, 20, 30, 60, 100, 120\rangle$ and from non-quantified compressed signal to study how quantification affects the signal reconstruction. The test result is shown in Fig. \ref{fig_prd}. We can see that PRD increases with the growth of CR, since compression can inevitably lose some information in the original signal. We can achieve at least "good" recoverability even when $qs = 100$, compressing signals by 90\%. It is clear that PRD grows with the increase of $qs$. This is because quantification makes the compressed signal $\mathbf y$ obtained by the receiver less accurate. When a smaller $qs$ is used (10 or 20), we can even gain a ``very good" signal recovery with $CR \leq 60$. The result shows we have a large feasibility in the tradeoff between reducing the communication overhead and ensuring the data precision.

\subsubsection{Secrecy in Data Transmission}
\label{subsec_attack_1}
The upper and lower bounds of ECG data used in our experiment are 1487 and 590, respectively, so we choose $d_i$ randomly from $[-449, 449]$. Without $\mathbf{K_d} = (d_1, \cdots, d_N)^{\mathbf T}$, COA attackers may attempt a uniform guess by assigning the same value $\bar{d}$ to all $d_i$ to derive the real compressed data $\mathbf y$. Then attackers can try different values for $\bar{d}$ and perform reconstruction algorithms to get the approximation $\hat{\mathbf x}$ of the original signal $\mathbf x$. To further study the effectiveness of this attack, we test the PRD of this attack under different $\bar{d}$ and $\rm CR$. The result is shown in Fig. \ref{fig_attack_d}.

From Fig. \ref{fig_attack_d}, the PRD minimizes when $\bar{d} = 0$, i.e., the best recovery this attack can achieve is executing reconstruction algorithm directly on the received $\mathbf{y'}$ or $\mathbf y''$.
Adding large enough random noise to the original signal has destroyed its sparsity, so the uniform guess attack has a poor performance.

We set up another experiment to compare the reconstruction quality by the legitimate receiver and the optimal uniform guess attacker under non-quantified and quantified situations. The result is shown in Fig. \ref{fig_attack_view} and \ref{fig_attack_qs}. Fig. \ref{fig_attack_view} gives a snapshot of the ECG record 230. The recovered signals deviate from the original signal in an irregular manner. In Fig. \ref{fig_attack_qs}, the normally recovered signals have a ``very good" quality, while the attacker only gets a bad recovery quality, no matter whether the encrypted signal is quantified. The results prove the secrecy provided by \emph{CSE}.


Besides the uniform guess, we also consider a random guess, where attackers guess values of $d_i$ randomly as how the smartphone generates. We repeat the random guess for 100 times and extract the PRDs under different $\rm CR$. The result in Fig. \ref{fig_attack_random} indicates the random guess still can only obtain a ``not good" quality.

\subsubsection{Overhead Analysis}

We focus on the computation and communication overhead on the IMD, since IMDs have very limited capacities and are hard to charge or replace. We test the execution time of different operations on our testbed, as shown in Table \ref{table_operation}. The execution time $T_{\rm CSEnc}$ of CS-based encryption relates tightly to the compression ratio (50, 75, 90 in Table \ref{table_operation}) which influences the size of $\mathbf{\Phi}$.

\begin{table}[!ht]
\small
\caption{Execution Time of Different Operations}
\label{table_operation}
\centering
\begin{tabular*}{0.95\linewidth}{@{\extracolsep{\fill}}c|c}
\toprule
Operation & Execution Time (ms)\\
\midrule
PBKDF2 \cite{kaliski2000pkcs} & 98.324\\
128-bit AES encryption & 0.736\\
128-bit AES decryption & 1.184\\
HMAC-SHA1 & 0.987 \\
CS-based encryption (CR=50,75,95) & 2.205,0.844,0.234, resp.\\
\bottomrule

\end{tabular*}
\vspace{-2mm}
\end{table}

Then we summarize the crypto-operations the IMD executes in each process, as shown in Table \ref{table_summary}. Thus, if $n$ denotes the number of 2-second ECG data to be transmitted in the read process, the IMD consumes 99.311, 3.158 + $n\cdot T_{\rm CSEnc}$, and 4.881 \emph{ms} for dealing with operations included in Table \ref{table_summary} in pairing, read, write processes, respectively. Note that the pairing process is only required the first time IMD was implanted, or the patient wants to pair another smartphone. IMDs attends other processes only when the patients come to see a doctor. Therefore, the extra computation overheads are acceptable for the user experience and IMD power consumption.

\begin{table}[!ht]
\small
\caption{Crypto-Operation Summary in Each Protocol}
\label{table_summary}
\centering
\begin{tabular*}{0.47\textwidth}{@{\extracolsep{\fill}}cccccc}
\toprule
Process        &    AES Enc & AES Dec & HMAC & KDF & CS Enc \\
\midrule
Pair        &    N/A     & N/A     & 1    &  1 &N/A\\
Auth &    N/A     & N/A     &N/A   & N/A & N/A\\
Read      &    N/A     & 1       & 2    & N/A & n\\
Write     &    1       & 1       & 3    & N/A  & N/A\\
\bottomrule
\end{tabular*}
\end{table}

From Fig. \ref{fig_prd}, we can conclude that our CS-based encryption can achieve a compression ratio $\rm CR=85$ when $qs=100$, and $\rm CR=95$ when $qs=60$ for ``good" quality, and $\rm CR=50$ when $qs=20$ and $\rm CR=60$ when $qs=10$ for ``very good" quality.
Therefore, our CS-based encryption scheme can reduce the communication overhead by over 50\%.

\section{Related Work}

\subsection{IMD Access Control}
Although various key management and access control schemes \cite{xiao2007survey, du2007effective, du2009routing, wu2014mobifish} have been proposed for building secure applications, few can be directly applied to secure IMD access. Some works use a pre-loaded key. Halperin \emph{et al.} \cite{halperin2008pacemakers} proposed to have all programmers carry a master key that can generate the IMD-specific key, using the device identity and a nonce provided by the IMD. However, forcing all IMD manufacturers to use the same master key is not realistic. The pre-loaded key can also be in the form of a rolling code \cite{HealthCom2011} or wake-up code \cite{Liu}. Some other works proposed to utilize the patient's biological features3 or physical objects in the patient's possession \cite{DenningCHI2010}, where the device key can be extracted from. However, the adversary could also find a way to illegally and secretly gain the bio-features or codes carried by the objects.

Another group of related works depends on symmetric temporary keys generated and/or distributed in real-time while the programmer is accessing the IMD in proximity. During key generation, the programmer and the IMD measure the same source, e.g. electrocardiogram signal (ECG) \cite{CCS2013}, interpulse interval \cite{poon2006novel}. The key distribution can be achieved via channels like body-coupled communications \cite{HealthSec2012}, vibration \cite{Vibration}, ultrasound \cite{rasmussen2009proximity}, and near field communication \cite{NFC2}.

Due to the limitations of IMD's computation and battery capacity, a proxy can be utilized to delegate the authentication on behalf of the IMD. Denning \emph{et al.} \cite{Cloaker} proposed to use an externally worn device \emph{Cloaker} to check whether an external programmer is authorized to communicate with the IMD, and to provide open access to all programmers by disabling the Cloaker. The Cloaker leverages heavyweight public key cryptography to verify the signature of the requesting programmer. Another scheme \emph{IMDGuard} is designed in \cite{xu2011imdguard}, in which a dedicated wearable device \emph{Guardian} is used to control the external programmer's access. The IMD and Guardian measure the patient's ECG signals simultaneously, from which a pair of shared keys is extracted to secure the link between the IMD and the Guardian. Therefore, an adversary cannot forge a fake Guardian except by making physical contact with the patient. 

Gollakota \emph{et al.} \cite{gollakota2011they} presented \emph{Shield}, which can defend against both eavesdropping and active attacks. Shield is a jammer-cum-receiver that receives the IMD's message while jamming others from decoding the message. Besides, Shield jams any other device's signal from communicating directly with the IMD, so that the adversary cannot trick the IMD into executing an unauthorized command. Zheng \emph{et al.} \cite{zheng2014non} proposed \emph{BodyDouble}, a non-key based scheme which employs an authentication proxy embedded in a gateway to authenticate a programmer. 

\vspace{-2mm}
\subsection{Compressive Sensing}
Compressive sensing \cite{donoho2006compressed, candes2006near} was first proposed for signal processing due to the feature that it unifies signal sensing and compression into signal acquisition and subsequently applied to various fields. Recently, compressive sensing has been extensively researched in the security and privacy domain, especially in scenarios where data need to be encrypted in resource-constrained devices while decrypted in entities with strong capabilities. Literature \cite{rachlin2008secrecy, orsdemir2008security} showed that CS-based encryption can achieve computational security when keeping the sensing matrix secret. Fragkiadakis et al. \cite{fragkiadakis2016enhancing} showed that the above CS-based encryption is vulnerable to Chosen Plaintext Attacks and proposed a scheme based on chaotic sequences to resist CPA attacks. Zhang et al. \cite{zhang2013energy} evaluated the performance of encompression (encryption+compression) on commercial embedded sensor platforms, showing that with a reasonable compression, encompression could considerably reduce the communication overhead and total energy consumption. Wang et al. \cite{wang2014privacy} formulated a privacy-preserving healthcare monitoring system for image data, employing compressive sensing to reduce the energy consumption of sensors by outsourcing the reconstruction to the cloud.

\vspace{-1mm}
\section{Conclusions}

A novel and practical system e-SAFE is presented to solve the access control and efficient transmission problems for IMDs. We enhance the practicality of our scheme by utilizing the most popular and handy device, smartphone, as a proxy to undertake most of the security-related tasks. The smartphone and IMD establish a trusted channel by sharing a secret key derived from a master key that is physically inaccessible to attackers. We authenticate the doctor who is responsible for all operations on patient's IMDs. We also give the keying material for authorizing access privileges to the doctor's programmer through the widely used SMS/email service. The patients and doctors are involved to perform critical operations manually for the authentication, reducing the chance that allows attackers to perform covert attacks. The patient's smartphone can keep evidence of the issued critical commands to his IMD for potential forensics. Experiment results based on our testbed demonstrate the effectiveness and efficiency of our scheme.

\bibliographystyle{IEEEtran}
\bibliography{reference}

\begin{thebibliography}{10}
\providecommand{\url}[1]{#1}
\csname url@samestyle\endcsname
\providecommand{\newblock}{\relax}
\providecommand{\bibinfo}[2]{#2}
\providecommand{\BIBentrySTDinterwordspacing}{\spaceskip=0pt\relax}
\providecommand{\BIBentryALTinterwordstretchfactor}{4}
\providecommand{\BIBentryALTinterwordspacing}{\spaceskip=\fontdimen2\font plus
\BIBentryALTinterwordstretchfactor\fontdimen3\font minus
  \fontdimen4\font\relax}
\providecommand{\BIBforeignlanguage}[2]{{%
\expandafter\ifx\csname l@#1\endcsname\relax
\typeout{** WARNING: IEEEtran.bst: No hyphenation pattern has been}%
\typeout{** loaded for the language `#1'. Using the pattern for}%
\typeout{** the default language instead.}%
\else
\language=\csname l@#1\endcsname
\fi
#2}}
\providecommand{\BIBdecl}{\relax}
\BIBdecl

\bibitem{rushanan2014sok}
M.~Rushanan, A.~D. Rubin, D.~F. Kune, and C.~M. Swanson, ``Sok: Security and
  privacy in implantable medical devices and body area networks,'' in
  \emph{IEEE Symposium on Security and Privacy 2014.}

\bibitem{Medtronic}
A.~Medtronic~Inc., ``Answers to questions about implantable cardiac devices,''
  \url{http://www.medtronic.com/content/dam/medtronic-com-m/mdt/documents/emc_guide.pdf},
  2016.

\bibitem{halperin2008pacemakers}
D.~Halperin, T.~S. Heydt-Benjamin, B.~Ransford, S.~S. Clark, B.~Defend,
  W.~Morgan, K.~Fu, T.~Kohno, and W.~H. Maisel, ``Pacemakers and implantable
  cardiac defibrillators: Software radio attacks and zero-power defenses,'' in
  \emph{IEEE Symposium on Security and Privacy 2008.}

\bibitem{HealthCom2011}
C.~Li, A.~Raghunathan, and N.~K. Jha, ``Hijacking an insulin pump: Security
  attacks and defenses for a diabetes therapy system,'' in \emph{IEEE Healthcom
  2011}.

\bibitem{kirk2012pacemaker}
J.~Kirk, ``Pacemaker hack can deliver deadly 830-volt jolt,''
  \emph{Computerworld}, vol.~17, 2012.

\bibitem{Liu}
J.~Liu, M.~A. Ameen, and K.~S. Kwak, ``Secure wake-up scheme for wbans,''
  \emph{{IEICE} Transactions on Communications}, vol. 93-B, no.~4, 2010.

\bibitem{CCS2013}
M.~Rostami, A.~Juels, and F.~Koushanfar, ``Heart-to-heart (h2h): authentication
  for implanted medical devices,'' in \emph{ACM CCS 2013}.

\bibitem{Vibration}
Y.~Kim, W.~S. Lee, V.~Raghunathan, N.~K. Jha, and A.~Raghunathan,
  ``Vibration-based secure side channel for medical devices,'' in
  \emph{ACM/EDAC/IEEE Design Automation Conference (DAC) 2015.}

\bibitem{rasmussen2009proximity}
K.~B. Rasmussen, C.~Castelluccia, T.~S. Heydt-Benjamin, and S.~Capkun,
  ``Proximity-based access control for implantable medical devices,'' in
  \emph{ACM CCS 2009}.

\bibitem{Cloaker}
T.~Denning, K.~Fu, and T.~Kohno, ``Absence makes the heart grow fonder: New
  directions for implantable medical device security,'' in \emph{HotSec 2008}.

\bibitem{xu2011imdguard}
F.~Xu, Z.~Qin, C.~C. Tan, B.~Wang, and Q.~Li, ``Imdguard: Securing implantable
  medical devices with the external wearable guardian,'' in \emph{IEEE INFOCOM
  2011.}

\bibitem{gollakota2011they}
S.~Gollakota, H.~Hassanieh, B.~Ransford, D.~Katabi, and K.~Fu, ``They can hear
  your heartbeats: non-invasive security for implantable medical devices,''
  \emph{ACM SIGCOMM Computer Communication Review}, vol.~41, no.~4, pp. 2--13,
  2011.

\bibitem{rostami2013balancing}
M.~Rostami, W.~Burleson, A.~Juels, and F.~Koushanfar, ``Balancing security and
  utility in medical devices?'' in \emph{ACM/EDAC/IEEE Design Automation
  Conference (DAC) 2013.}

\bibitem{poh2011advancements}
M.-Z. Poh, D.~J. McDuff, and R.~W. Picard, ``Advancements in noncontact,
  multiparameter physiological measurements using a webcam,'' \emph{IEEE
  transactions on biomedical engineering}, vol.~58, no.~1, pp. 7--11, 2011.

\bibitem{cheng2017lightweight}
Y.~Cheng, X.~Fu, X.~Du, B.~Luo, and M.~Guizani, ``A lightweight live memory
  forensic approach based on hardware virtualization,'' \emph{Information
  Sciences}, vol. 379, pp. 23--41, 2017.

\bibitem{wu2014security}
L.~Wu, X.~Du, and X.~Fu, ``Security threats to mobile multimedia applications:
  Camera-based attacks on mobile phones,'' \emph{IEEE Communications Magazine},
  vol.~52, no.~3, pp. 80--87, 2014.

\bibitem{liang2014permission}
S.~Liang and X.~Du, ``Permission-combination-based scheme for android mobile
  malware detection,'' in \emph{IEEE ICC 2014}.

\bibitem{du2008security}
X.~Du and H.-H. Chen, ``Security in wireless sensor networks,'' \emph{IEEE
  Wireless Communications}, vol.~15, no.~4, 2008.

\bibitem{donoho2006compressed}
D.~L. Donoho, ``Compressed sensing,'' \emph{IEEE Transactions on information
  theory}, vol.~52, no.~4, pp. 1289--1306, 2006.

\bibitem{candes2006near}
E.~J. Candes and T.~Tao, ``Near-optimal signal recovery from random
  projections: Universal encoding strategies?'' \emph{IEEE transactions on
  information theory}, vol.~52, no.~12, pp. 5406--5425, 2006.

\bibitem{tropp2007signal}
J.~A. Tropp and A.~C. Gilbert, ``Signal recovery from random measurements via
  orthogonal matching pursuit,'' \emph{IEEE Transactions on information
  theory}, vol.~53, no.~12, pp. 4655--4666, 2007.

\bibitem{figueiredo2007gradient}
M.~A. Figueiredo, R.~D. Nowak, and S.~J. Wright, ``Gradient projection for
  sparse reconstruction: Application to compressed sensing and other inverse
  problems,'' \emph{IEEE Journal of selected topics in signal processing},
  vol.~1, no.~4, pp. 586--597, 2007.

\bibitem{orsdemir2008security}
A.~Orsdemir, H.~O. Altun, G.~Sharma, and M.~F. Bocko, ``On the security and
  robustness of encryption via compressed sensing,'' in \emph{IEEE MILCOM
  2008}.

\bibitem{rachlin2008secrecy}
Y.~Rachlin and D.~Baron, ``The secrecy of compressed sensing measurements,'' in
  \emph{IEEE Annual Allerton Conference on Communication, Control, and
  Computing 2008}.

\bibitem{kaliski2000pkcs}
B.~Kaliski, ``Pkcs\# 5: Password-based cryptography specification version
  2.0,'' 2000.

\bibitem{arias2011noise}
E.~Arias-Castro and Y.~C. Eldar, ``Noise folding in compressed sensing,''
  \emph{IEEE Signal Processing Letters}, vol.~18, no.~8, pp. 478--481, 2011.

\bibitem{wang2017recovery}
B.~Wang, L.~Hu, J.~An, G.~Liu, and J.~Cao, ``Recovery error analysis of noisy
  measurement in compressed sensing,'' \emph{Circuits, Systems, and Signal
  Processing}, vol.~36, no.~1, pp. 137--155, 2017.

\bibitem{goldberger2000physiobank}
A.~L. Goldberger, L.~A. Amaral, and et~al., ``Physiobank, physiotoolkit, and
  physionet,'' \emph{Circulation}, vol. 101, no.~23, pp. e215--e220, 2000.

\bibitem{jalaleddine1990ecg}
S.~M. Jalaleddine, C.~G. Hutchens, R.~D. Strattan, and W.~A. Coberly, ``Ecg
  data compression techniques-a unified approach,'' \emph{IEEE transactions on
  Biomedical Engineering}, vol.~37, no.~4, pp. 329--343, 1990.

\bibitem{zigel2000weighted}
Y.~Zigel, A.~Cohen, and A.~Katz, ``The weighted diagnostic distortion (wdd)
  measure for ecg signal compression,'' \emph{IEEE Transactions on Biomedical
  Engineering}, vol.~47, no.~11, pp. 1422--1430, 2000.

\bibitem{xiao2007survey}
Y.~Xiao, V.~K. Rayi, B.~Sun, X.~Du, F.~Hu, and M.~Galloway, ``A survey of key
  management schemes in wireless sensor networks,'' \emph{Computer
  communications}, vol.~30, no. 11-12, pp. 2314--2341, 2007.

\bibitem{du2007effective}
X.~Du, Y.~Xiao, M.~Guizani, and H.-H. Chen, ``An effective key management
  scheme for heterogeneous sensor networks,'' \emph{Ad Hoc Networks}, vol.~5,
  no.~1, pp. 24--34, 2007.

\bibitem{du2009routing}
X.~Du, M.~Guizani, Y.~Xiao, and H.-H. Chen, ``A routing-driven elliptic curve
  cryptography based key management scheme for heterogeneous sensor networks,''
  \emph{IEEE Transactions on Wireless Communications}, vol.~8, no.~3, pp.
  1223--1229, 2009.

\bibitem{wu2014mobifish}
L.~Wu, X.~Du, and J.~Wu, ``Mobifish: A lightweight anti-phishing scheme for
  mobile phones,'' in \emph{IEEE ICCCN 2014}.

\bibitem{DenningCHI2010}
T.~Denning, A.~Borning, B.~Friedman, B.~T. Gill, T.~Kohno, and W.~H. Maisel,
  ``Patients, pacemakers, and implantable defibrillators: Human values and
  security for wireless implantable medical devices,'' in \emph{SIGCHI
  Conference on Human Factors in Computing Systems}, 2010.

\bibitem{poon2006novel}
C.~C. Poon, Y.-T. Zhang, and S.-D. Bao, ``A novel biometrics method to secure
  wireless body area sensor networks for telemedicine and m-health,''
  \emph{IEEE Communications Magazine}, vol.~44, no.~4, pp. 73--81, 2006.

\bibitem{HealthSec2012}
S.-Y. Chang, Y.-C. Hu, H.~Anderson, T.~Fu, and E.~Y. Huang, ``Body area network
  security: Robust key establishment using human body channel.'' in
  \emph{HealthSec}, 2012, pp. 5--5.

\bibitem{NFC2}
B.~Kim, J.~Yu, and H.~Kim, ``In-vivo nfc: Remote monitoring of implanted
  medical devices with improved privacy,'' in \emph{ACM SenSys 2012}.

\bibitem{zheng2014non}
G.~Zheng, G.~Fang, M.~A. Orgun, and R.~Shankaran, ``A non-key based security
  scheme supporting emergency treatment of wireless implants,'' in \emph{IEEE
  ICC 2014}.

\bibitem{fragkiadakis2016enhancing}
A.~Fragkiadakis, L.~Kovacevic, and E.~Tragos, ``Enhancing compressive sensing
  encryption in constrained devices using chaotic sequences,'' in \emph{ACM
  SmartObjects 2016}.

\bibitem{zhang2013energy}
M.~Zhang, M.~M. Kermani, A.~Raghunathan, and N.~K. Jha, ``Energy-efficient and
  secure sensor data transmission using encompression,'' in \emph{IEEE VLSID
  2013}.

\bibitem{wang2014privacy}
C.~Wang, B.~Zhang, K.~Ren, J.~M. Roveda, C.~W. Chen, and Z.~Xu, ``A
  privacy-aware cloud-assisted healthcare monitoring system via compressive
  sensing,'' in \emph{IEEE INFOCOM 2014}.

\end{thebibliography}

%
%

\end{document}